%% file: paper.tex
\documentclass[12pt]{iopart}
\usepackage[english]{babel}
\usepackage[utf8]{inputenc}
\usepackage{xcolor}
\usepackage{graphicx}
\usepackage{float}
\usepackage{hyperref}
\usepackage[margin=2.0cm,bottom=3cm]{geometry}
\usepackage{paralist}
\usepackage{a4}
\usepackage{tabu}
\usepackage{dsfont}
\usepackage{siunitx}
\usepackage{amsthm, amssymb}

\newcommand{\eq}[1]{$\mathrm{Eq.}$~\ref{#1}}

\newcommand{\figref}[1]{Fig.~\ref{#1}}

\newcommand{\tabref}[1]{Tab.~\ref{#1}}

\newcommand{\fricU}[1]{\ensuremath{\SI[per-mode=symbol]{#1}{\kg\per\s}}}
\newcommand{\um}[1]{\ensuremath{\SI{#1}{\micro\meter}}}

\newcommand{\s}[1]{\ensuremath{\SI[per-mode=symbol]{#1}{\s}}}

\newcommand{\velU}[1]{\ensuremath{\SI[per-mode=symbol]{#1}{\micro\meter\per\s}}}

\usepackage{iopams}  
\begin{document}

\title[Active Poroelastic Two-Phase Model]{Oscillatory motion of a droplet in an active poroelastic two-phase model}

\author{Dirk Alexander Kulawiak\textsuperscript{1}, Jakob Löber\textsuperscript{2}, Markus Bär\textsuperscript{3}, and Harald Engel\textsuperscript{1}}

\address{\textsuperscript{1} TU Berlin - Institut für Theoretische Physik, Hardenberstr. 36, 10623 Berlin, Germany; \textsuperscript{2} Max-Planck-Institut für Physik komplexer Systeme, Nöthnitzer Straße 38, 01187 Dresden, Germany; \textsuperscript{3} Physikalisch-Technische Bundesanstalt, Abbestrasse 2-12, 10587 Berlin, Germany}
\ead{harald.engel@tu-berlin.de}
\vspace{10pt}
\begin{indented}
\item[]July 2018
\end{indented}

\begin{abstract}
The onset of self-organized droplet motion is studied in a poroelastic two-phase model with free boundaries and substrate friction. In the model, an active, gel-like phase and a passive, fluid-like phase interpenetrate on small length scales. A feedback loop between a chemical regulator, mechanical deformations, and induced fluid flow gives rise to oscillatory and irregular droplet motion accompanied by spatio-temporal contraction patterns inside the droplet. By numerical simulations in one spatial dimension, we cover extended parameter regimes of active tension and substrate friction, and reproduce experimentally observed oscillation periods and amplitudes. In line with recent experiments, the model predicts alternating forward and backward fluid flow at the boundaries with reversed flow in the center. Our model is a first step towards a more detailed model of moving microplasmodia of Physarum polycephalum.
\end{abstract}

%
%
%
%
%

\input{introduction}
\input{model}
\input{results}
\input{discussion}
\input{method}

\section*{Acknowledgments}
We thank Markus Radszuweit for helpful discussions about his previous work on the model. Computational resources were provided by the Institut für Theoretische Physik at the TU Berlin. DAK was funded by the German Science Foundation (DFG) within the GRK 1558.

\section{References}
\bibliographystyle{iopart-num}
\bibliography{paper}{}

\end{document}

%% file: introduction.tex
\section{Introduction}
Dynamic processes in cells, and cell motility in particular, are intriguing examples of large-scale 
spatio-temporal order in systems far from thermodynamic equilibrium \cite{murray_mathematical_2007,winfree_geometry_2001,prigogine_time_1978}.
Here, the continuous turnover of ATP by molecular motors \cite{kolomeisky_molecular_2007} provides the energy to drive mechano-chemical contraction-expansion patterns and, ultimately, locomotion. Biological examples of these phenomena are reviewed and discussed in \cite{gross_how_2017,nishikawa_controlling_2017,howard_turings_2011,mayer_anisotropies_2010,kumar_pulsatory_2014}.

A well-studied model organism exhibiting a huge variety of spatio-temporal mechano-chemical patterns with and without 
locomotion is the true slime-mold \textit{Physarum polycephalum} \cite{oettmeier_physarum_2017,aldrich_cell_1982,teplov_role_2017}. 
Physarum is unicellular, but a cell contains multiple nuclei and can grow to the size of several 
square meters \cite{fessel_structuring_2015}. Physarum microplasmodia are an artificial form of Physarum with a size between \um{100} and \SI{1}{\milli\meter} that do not occur in nature \cite{takagi_emergence_2008,takagi_annihilation_2010}. They are composed of a gel-like ectoplasm and a fluid-like endoplasm \cite{zhang_self-organized_2017,oettmeier_form_2018}. Microplasmodia are produced by extracting
a  sufficient  amount  of cytoplasm from a Physarum cell and placing it on a substrate. After reorganization,
such a protoplasmic droplet displays a wide variety of mechano-chemical patterns such as standing, 
traveling, and spiral waves as well as irregular oscillations \cite{takagi_emergence_2008}. 
After several hours, a cell elongates into a tadpole-like shape and starts to explore its surroundings 
\cite{rieu_periodic_2015,rodiek_migratory_2015,rodiek_patterns_2015,lewis_coordination_2015}.

Movement of microplasmodia occurs in two modes: \textit{peristaltic} and \textit{amphistaltic} \cite{zhang_self-organized_2017, lewis_coordination_2015}. In both modes, microplasmodia alternate between forward and backward motion with a well-defined period. The forward motion is larger than the backward motion, resulting in a net displacement within each period. In the more frequently observed peristaltic mode, motion is driven by mechano-chemical waves originating at the tail and traveling towards the front. In the amphistaltic mode, front and tail contract in anti-phase oscillations.

Common models for the cytoskeleton are based on active fluid and gel models \cite{julicher_active_2007,joanny_active_2009,kopf_non-equilibrium_2013,bois_pattern_2011, strychalski_poroelastic_2015}. In contrast, some models for the crawling type of amoeboid cell motility \cite{kulawiak_modeling_2016,lober_modeling_2014} neglect intracellular
flows. As opposed to simple fluids and solids, which are governed by a single momentum balance equation, 
poroelastic media belong to the class of two-fluid models. These are characterized by individual momentum balance equations for each of 
the constitutive phases. Such a description is useful if two phases with largely different rheological properties interpenetrate on 
relatively small length scales, such as groundwater permeating porous rock \cite{coussy_poromechanics_2004}, the superposition 
of normal and inviscid superfluid helium in helium II \cite{landau_fluid_1987}, or cytosol pervading the cytoskeleton.

Poroelastic two-phase models have been used successfully \cite{radszuweit_active_2014,radszuweit_model_2010,alonso_oscillations_2016} as ingredients in detailed models to replicate the pattern found in resting microplasmodia of Physarum \cite{takagi_emergence_2008}.
Our work is based on the simple generic model of a poroelastic active droplet introduced by Radszuweit et al. in \cite{radszuweit_intracellular_2013}.
Therein, a feedback loop between an advected chemical regulator and active stress gives rise to self-organized spatio-temporal contraction patterns. This occurs even without the inclusion of a nonlinear reaction-diffusion kinetics for the chemical regulator, that were part of the detailed Physarum models mentioned above.

Appropriate boundary conditions must be introduced to close these poroelastic models and all earlier approaches utilized fixed boundaries to study resting microplasmodia.
While these fixed boundaries are simpler to implement in numerical simulations and allow us to study mechanical deformations in the bulk, they preclude the possibility of deformations of the droplet boundary and therefore motion of the droplet as a whole.

Here, we introduce free boundary conditions that allow for deformations and motion of the droplet boundary.
Considering a free boundary problem complicates numerical simulations and can significantly change the solution of a given problem, especially for fluid dynamics \cite{balossino_uence_2006,friedman_free_2015}.

Previous work has shown that our model is able to exhibit self-organized spatially non-symmetric deformations \cite{radszuweit_intracellular_2013}. In order to describe the impact of spatio-temporal deformation patterns on the motion of the now free boundaries, we must include a substrate friction into the model. The value of the friction coefficient strongly affects the resulting motion of the active poroelastic two-phase droplet in our minimal model. The aim of this work is to explore the conditions for the motion of a droplet in a minimum model of an active poroelastic medium. While we are motivated by the observations of Physarum microplasmodia, we do not intend to provide a realistic model for the latter.

Section 2 contains a comprehensive description of our model. Section 3 gives results on the relation of the possibility of the motion of the droplet with the spatio-temporal mechano-chemical patterns and their symmetries. Furthermore, we show a phase diagram of droplet motility and occurrence of patterns as a function of the mechano-chemical feedback strength (defining the strength of the active tension) and the friction coefficient (quantifying the strength of friction of the droplet with the substrate). Moreover, Section 3 contains a comparison of  the findings in the simple model of a moving active poroeleatic droplet with recent experiments in Physarum microplasmodia. The discussion explains why our model cannot show net motion and briefly points towards future work addressing a complete model for the motion of Physarum microplasmodia.

%% file: model.tex
\section{Model}
In our poroelastic two-phase model, the homogeneous isotropic droplet consists of an active, gel-like phase and a passive, fluid-like phase that interpenetrate at relatively small length scales \cite{alt_cytoplasm_1999,joanny_hydrodynamic_2007}. The passive phase flows with velocity v. The active gel phase is a visco-elastic solid with mechanical displacements u and velocity $\dot{u}$.

Both phases individually satisfy a momentum balance equation expressed with stress tensors, where $\sigma_g$ and $\sigma_f$ denote the stress in the gel and in the fluid, respectively. The total stress is given by $\sigma = \rho_g\sigma_g + \rho_f\sigma_f$, with $\rho_g$ ($\rho_f$) denoting the volume fraction of the gel (fluid) phase. Deformations of the medium result in an exchange of volume between the respective fractions. The time evolution for the gel fraction is given by $ \hat{\rho}_g = \rho_g \left( 1 - \partial_x u \right)$. Here, $\rho_g$ is the initial, spatially constant gel fraction. However, due to our small strains approximation, only the constant term $\rho_g$ enters in our model equations. Furthermore, we assume that there are no other phases present, and the volume fractions satisfy $\rho_g + \rho_f = 1$ at all times \cite{dembo_cell_1986}. 

A droplet occupies a one dimensional, time dependent domain $\mathcal{B}$ with boundaries denoted by $\partial \mathcal{B}$.
Assuming that $\mathcal{B}$ is infinitely large in the y-direction, the boundary is straight, and we omit terms that depend on interface tension or bending.
Free boundary conditions in x-direction enable the boundary to deform and move in response to bulk flow and deformation \cite{larripa_transport_2006,nickaeen_free-boundary_2017,kopf_continuum_2013}. 
We assume that the droplet is surrounded by an inviscid fluid described with stress tensor $\sigma_\mathrm{out} = -p_\mathrm{out}$. The exact value of the outside hydrostatic pressure $p_\mathrm{out}$ is not important, as long as it is constant and homogeneous. Thus, we choose $p_\mathrm{out} = 0$.
At the droplet's boundary, the total stress has to be continuous across the interface. This gives the first boundary condition
\begin{eqnarray}\label{eq_BCstress}
\sigma - p\Big{|}_{\partial \mathcal{B}} =  \sigma_\mathrm{out}\Big{|}_{\partial \mathcal{B}}  = 0,
\end{eqnarray}
where the subscript $\partial \mathcal{B}$ denotes evaluation at the boundary of domain $\mathcal{B}$. Because of the two momentum balance relations in the poroelastic model, we need a second boundary condition. Assuming there is no polymerization of actin at and no permeation of the fluid phase through the boundary, the velocity of gel and fluid must match. This gives rise to the additional boundary condition
\begin{eqnarray}\label{eq_unique}
\dot{u}\Big{|}_{\partial \mathcal{B}} = v\Big{|}_{\partial \mathcal{B}}.
\end{eqnarray}
Note, that the free boundary conditions require the evaluation of stress tensors and flow fields at the boundary, whose position itself must be determined in the course of solving the evolution equations. We circumvent this problem by transforming the system to a co-moving frame of reference.
In general, continuum mechanics allows us to use different coordinate frames to formulate the model equations \cite{chaves_notes_2013}. 

We distinguish between the lab frame (LF) with \textit{spatial} coordinates $X$ and the gel's body reference frame (BRF) with \textit{material} coordinates $x$. The material displacement field $u$ connects both frames by $u(x,t) = X(x,t) - x$. Note that the domain as well as its boundary, which is time-dependent in the LF, becomes stationary in the BRF. The gel velocity $\dot{u}$ is given by the material time derivative $\dot{u} = \partial_tu +(\partial_x u) \dot{x} $. By definition, the gel is fixed in its BRF $(\dot{x} = 0)$, and the material time derivative simplifies to $\dot{u} = \partial_t u$. On the downside, transforming stress tensors given by linear constitutive laws from the LF to the BRF gives rise to many geometric nonlinearities. We simplify by linearizing in the strains, i.e., assuming $|\partial_x u| \ll 1$. However, note that we do not assume the displacements $u$ to be small. The displacements may, for example, grow linearly in time without bounds for a droplet moving with constant center of mass velocity. See \cite{radszuweit_intracellular_2013,radszuweit_active_2014} for details on the transformation from the LF to the BRF and \figref{fig_BRF} for a visual comparison of a quantity plotted in the BRF and the LF. 

The fluid phase is modeled as a passive viscous liquid with stress $\sigma_f = \eta_f\partial_x v$, where $\eta_f$ is the viscosity. The stress $\sigma_g = \sigma^\mathrm{ve} + \sigma^\mathrm{act}$ of the gel phase is decomposed in a passive part, $\sigma^\mathrm{ve}$, and an active part, $\sigma^\mathrm{act}$. We assume that the passive part is a viscoelastic Kelvin-Voigt solid with $\sigma^\mathrm{ve} = E \partial_x u + \eta_g \partial_x \dot{u}$ \cite{banks_brief_2011}. Here, $E$ is Young's modulus and $\eta_g$ the viscosity. Recently, the effects of alternative viscoelastic models for the passive gel stress were investigated in \cite{alonso_mechanochemical_2017}. A poroelastic model with nonlinear elasticity was introduced in \cite{taber_poroelastic_2011}.

Hence, we can write momentum balances in the BRF for both phases. The Reynolds numbers that arise from flows in the medium are assumed to be small and $\mathrm{Re} \ll 1$. Thus, inertia effects can be neglected, and the intra-droplet flow is described by the Stokes equation. The momentum balances read
\begin{eqnarray}
\rho_g\partial_x\left(\sigma_g-p\right) + f_g + f_\mathrm{fric} &= 0,\label{eq_ForceBal_1}\\
\rho_f\partial_x\left(\sigma_f-p\right)  + f_f &= 0,\label{eq_ForceBal_2}
\end{eqnarray}
where $p$ is the pressure stemming from the incompressibility of the medium expressed as\begin{equation}\label{eq_inc}
\partial_x\left(\rho_g\dot{u} + \rho_f v\right) = 0.
\end{equation}
The friction between both phases is given by Darcy’s law together with Newton's third law $f_g = -f_f = \rho_g\rho_f\beta\left(v-\dot{u}\right)$. We assume a linear substrate friction force $f_\mathrm{fric}= -\rho_g\gamma\dot{u}$ for friction between gel and substrate, and no friction between fluid and substrate.

The active stress is assumed to be governed by the concentration $c$ of a chemical regulator species
\begin{equation}\label{eq_act}
	\sigma^\mathrm{act} = T(c)= T_0 - \xi\frac{c}{1+c}.
\end{equation}
Here, $T_0$ is a homogeneous stress that is inhibited by the regulator $c$, and $\xi$ describes the strength of this active stress. This dependency is in contrast with the assumption of an activating regulator species in the simple model of a one-component active fluid \cite{bois_pattern_2011} and in line with observation regarding the effect of calcium in Physarum microplasmodia 
\cite{yoshimoto_simultaneous_1981}. 

The regulator $c$ is dissolved in the fluid and advected with the fluid flow $v$ in the LF. 
Furthermore, the regulator is diffusing with a diffusion coefficient $D_c$. Transforming the advection-diffusion equation from the LF to the BRF, and 
linearizing in the gel strains $\partial_x u$, yields an advection-diffusion equation with the relative velocity of the fluid to the gel  $v - \dot{u}$ as the advection velocity,
\begin{eqnarray}\label{eq_cal}
\partial_t c + \partial_x\left[\left(v - \dot{u}\right)c\right] = D_c\partial_{xx} c.
\end{eqnarray}
We assume that no regulator molecules can cross the droplet's membrane, resulting in a no-flux boundary condition for the variable $c$. Thus, the total amount of the regulator is conserved. Nevertheless, there can be local differences in the regulator concentration, which in turn drive mechanical deformations via a spatially varying active stress $T(c)$. When necessary, a nonlinear reaction kinetics for the regulator species can be taken into account, for details see \cite{radszuweit_active_2014,alonso_oscillations_2016}.

\noindent
In summary, the model equations are given by
\begin{eqnarray}
\rho_f\eta_f\partial_{xx} v + \rho_g\eta_g\partial_{xx} \dot{u} + \rho_g E\partial_{xx} u \label{eq_FullEq_a}
\nonumber\\
- \rho_g\gamma \dot{u} - \partial_x p &=  -\rho_g\partial_x T(c)\\
\eta_f\partial_{xx} v - \rho_g\beta(v-\dot{u}) - \partial_x p &=0\\
\partial_x (\rho_g \dot{u} + \rho_f v) &= 0	\\
\partial_t c + \partial_x\left[\left(v - \dot{u}\right)c\right] - D_c\partial_{xx} c&= 0.\label{eq_FullEq_d}
\end{eqnarray}
Note, that all patterns in our model emerge through self-organization. Furthermore, the only nonlinear terms are the advection term for the regulator and the active tension term $\partial_x T(c)$. We introduce the dimensionless P\'eclet number $\mathrm{Pe} = \xi/(D_c\beta)$ as a measure for the ratio of diffusive to advective time scales to characterize the strength of the active tension \cite{radszuweit_intracellular_2013}.

%% file: results.tex
\section{Results}
Equations \ref{eq_FullEq_a}-\ref{eq_FullEq_d} are solved with parameter values adopted from \cite{radszuweit_intracellular_2013} and listed in \tabref{tab_param} unless stated otherwise. The initial condition is the weakly perturbed homogeneous steady state (HSS) with $u =\dot{u} =v = 0$ and $c=c_0 = 1$. Due to the incompressibility of the medium (\eq{eq_inc}), the droplet's length $L$ is constant, and the location of the droplet's left boundary is used as a measure of its position. We observe fluid flow coupled to concentration and deformation patters both for resting and moving boundaries. Depending on parameter values of the friction coefficient $\gamma$ and the P\'eclet number $\mathrm{Pe}$, the droplet's position over time remains fixed or undergoes regular or irregular oscillations. However, in all cases, the temporal average of the droplet's position vanishes, i.e. there is no net motion. To visualize the dynamics of regulator and gel flow, we show space-time plots of these quantities in the BRF.

\begin{figure}[!ht]
	\centering
	\includegraphics[width=0.495\textwidth]{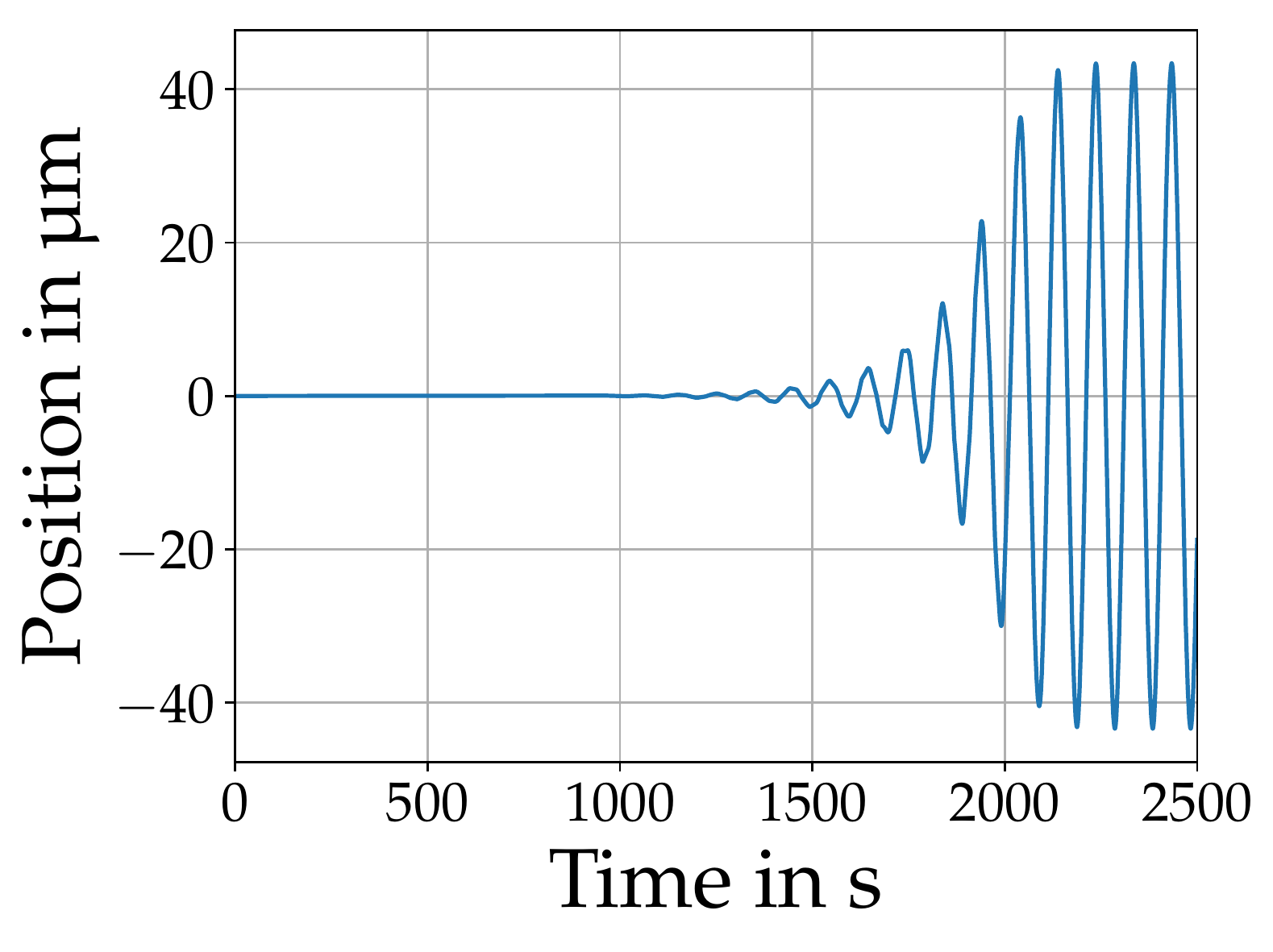}
	\includegraphics[width=0.495\textwidth]{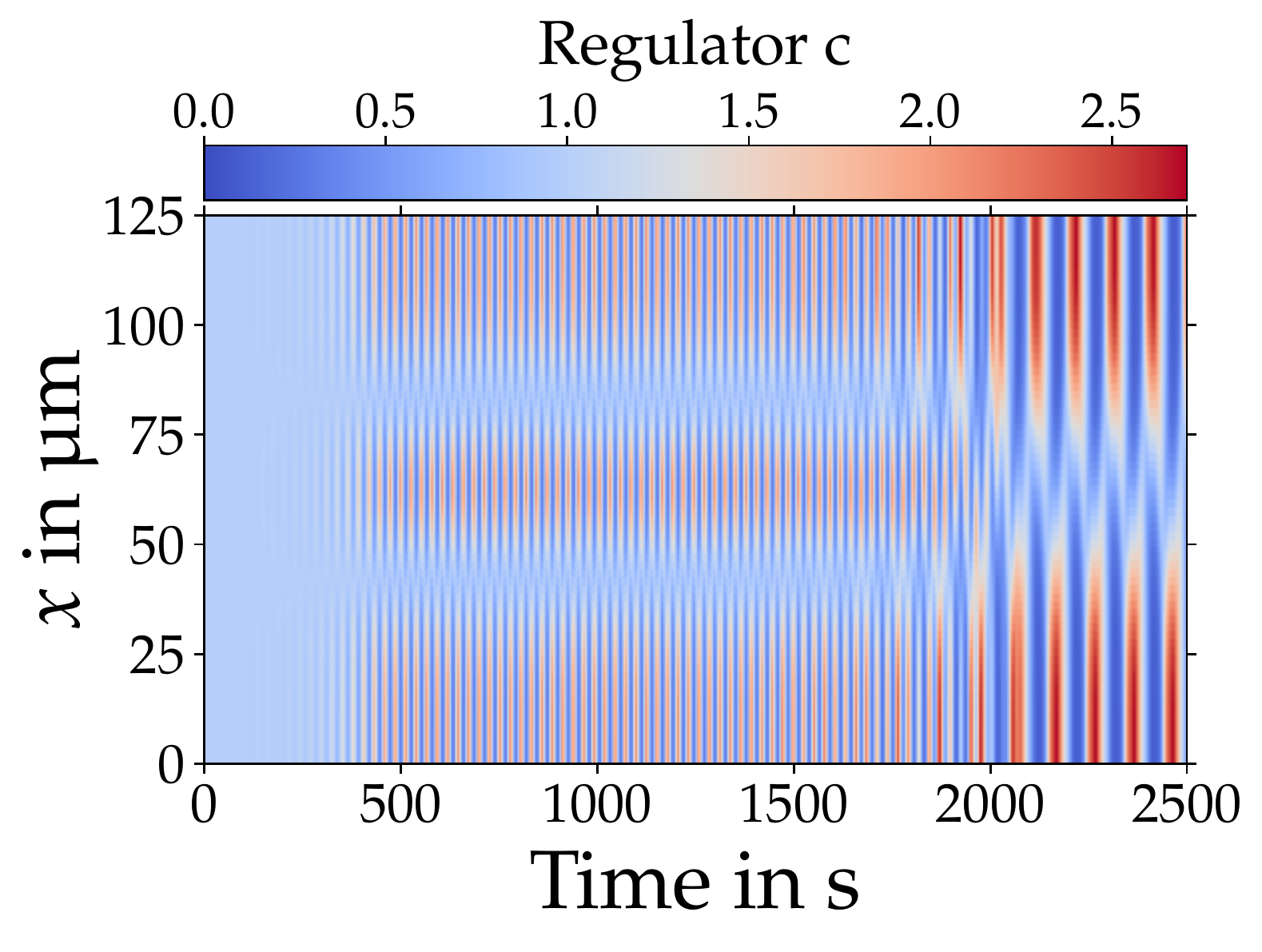}
	\caption{\textbf{Regular oscillations of the droplet's position over time (top) are accompanied by spatially antisymmetric regulator oscillations (bottom).} The active tension ($\mathrm{Pe} = 6$) is slightly above the critical value where the HSS destabilizes. After initialization, the resting droplet exhibits spatially symmetric regulator oscillations with period $T_\mathrm{s} = \s{24}$ and slowly growing amplitude. After $\approx \s{2000}$ a transition to spatially antisymmetric regulator oscillations occurs. Simultaneously, the droplet's position starts to oscillate with an amplitude of about 1/3 of its length and a period $T_\mathrm{as} = \s{98}$. The regulator concentration is plotted in a body reference frame co-moving with the gel phase.}
	\label{fig_xi60}
\end{figure}

\begin{figure*}[!ht]
	\centering
	\includegraphics[width=\textwidth]{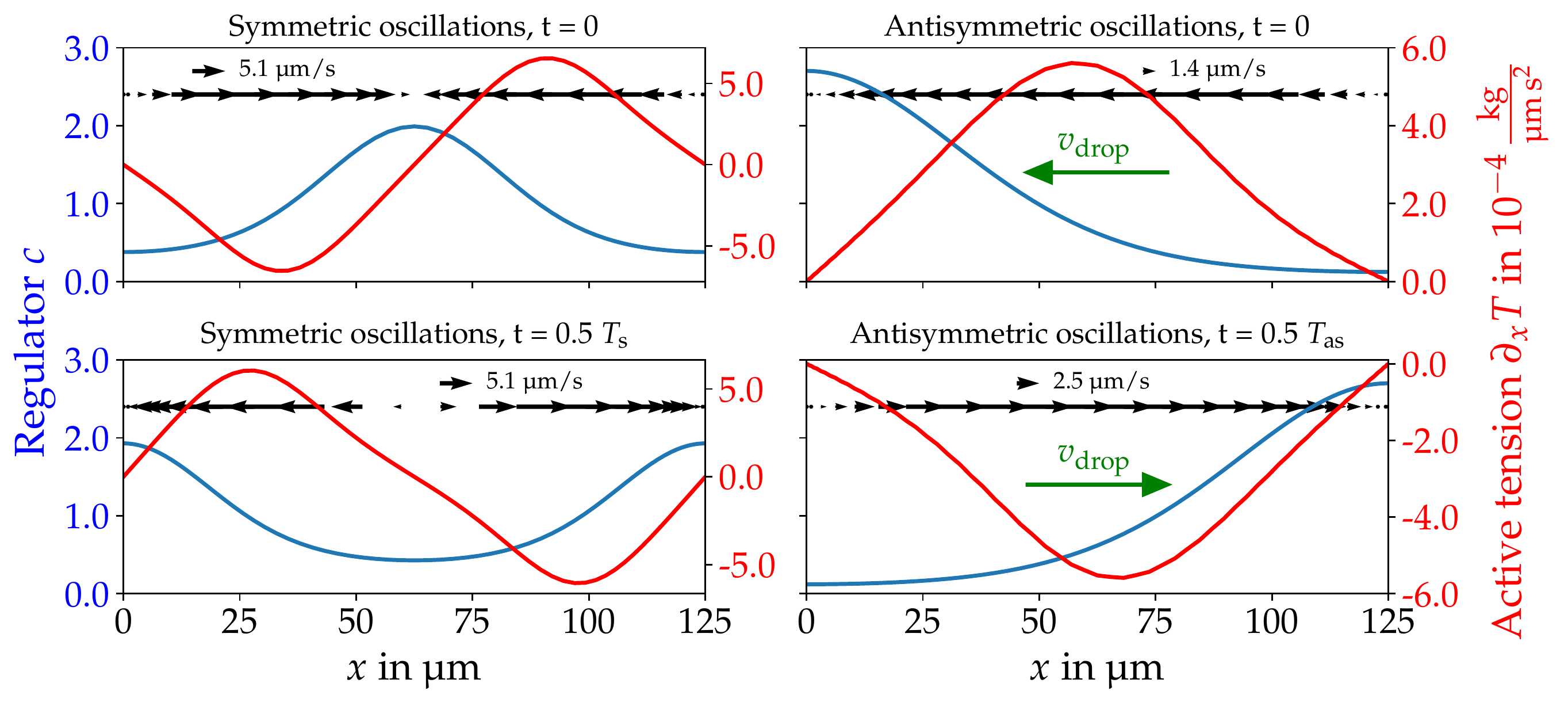}
	\caption{\textbf{Snapshots of spatially symmetric (left)  and antisymmetric (right) regulator oscillations.} The regulator concentration $c$ is depicted in blue, the active tension $\partial_x T \sim \partial_x c/(1+c)$ in red, and black arrows indicate the advection velocity $v-\dot{u}$. The length of the black arrows indicates the amplitude of the advective flow.
	Spatially symmetric oscillations have a period of $T_\mathrm{s} = \s{24}$ and $c$  oscillates between high concentration at the center (top left) and high concentration at the boundaries (bottom left). No movement occurs.
	In the second case, $c$ oscillates between two configurations antisymmetric to each other with a period of $T_\mathrm{as} = \s{98}$ (right). The droplet is moving with velocity $v_\mathrm{drop}$ towards the direction of high regulator concentration as indicated by the green arrows.
}
	\label{fig_xi60Snap}
\end{figure*}

\subsection*{Symmetric and antisymmetric spatio-temporal oscillations}\label{sec_SymOsc}

The case of a regular oscillation in \figref{fig_xi60} exemplifies how the different parts of the model interact and motion arises. Shortly after initialization, the concentration oscillates in a spatially symmetric manner with period $T_\mathrm{s} = \s{24}$ (\figref{fig_xi60}, bottom) and a constant position of the droplet (\figref{fig_xi60}, top). The blue lines in \figref{fig_xi60Snap} show snapshots of the regulator concentration as a function of space. The regulator distribution changes periodically from a high concentration at the center and low values at the boundaries (top left panel) to a low concentration at the center and high values at the boundaries (bottom left panel). In the top left panel, the active tension $\partial_x T \sim \partial_x c/(1+c)$ (red line, \eq{eq_act}) generates a symmetric advection flow from the boundaries towards the center (black arrows). This results in even more regulator piling up at the center, thus causing an even stronger flow. The flow deforms the droplet and elastic tension in the gel builds up. Over time, this elastic tension increases and counteracts to the active tension. The advection of regulator weakens, and at some point regulator diffusion takes over. The regulator concentration at the center starts to decrease, diminishing active tension and thereby advection even further. When the elastic tension overcomes the active tension the direction of advection changes. The concentration starts to pile up at the boundaries (bottom left panel), and the process repeats. During the whole oscillation cycle both the profiles of gel $\dot{u}$ and fluid $v$ are symmetric in space. Hence, spatially symmetric oscillations result in immobile droplets.

At $t \approx\s{1800}$ in \figref{fig_xi60} a transition from symmetric to antisymmetric regulator oscillations occurs. Here, the regulator concentration is high at one boundary but low at the other. With increasing amplitude of the antisymmetric oscillation, the droplet starts to move periodically back and forth. After a transition period of $\approx\s{200}$, the amplitude of regulator oscillations saturates, and the droplet performs periodic motion in phase with the regulator oscillations and a period of $T_\mathrm{as}=\s{98}$. \figref{fig_xi60Snap} (right) shows two snapshots of antisymmetric oscillations at different times. When the droplet's position is at its maximum displacement to the right, the regulator concentration changes from having its maximum at the right boundary to being higher on the left boundary.
When the maximum of the concentration switches sides, the active tension changes sign, causing an advective flow to the left. A rapid movement to the left starts with a maximum droplet velocity of $v_\mathrm{drop} = \velU{2.47}$. While this movement takes place, the concentration piles up on the left boundary (top right) and the elastic tension inside the droplet builds up.
Similar to the symmetric case, this elastic tension is acting in opposition to the active tension yielding a lower advection and droplet velocity. At some point, advection becomes weaker than diffusion. Then, the concentration has reached its maximum and begins to decrease. While the active tension is diminishing, the elastic stress inside the gel phase still builds up. Once it overcomes the active tension, the tension in the droplet starts to decay. This causes a change of direction of the advective flow, leading to a rise of concentration on the right, and the process repeats (bottom right). This antisymmetric oscillation results in a periodic movement with a maximum displacement of $u_\mathrm{max}\approx\um{42}$ but not in net motion.

\subsection*{Varying active tension strength and substrate friction excites different modes of motion}\label{sec_ActT}
\begin{figure}[H]
	\centering
	\includegraphics[width=0.44\linewidth]{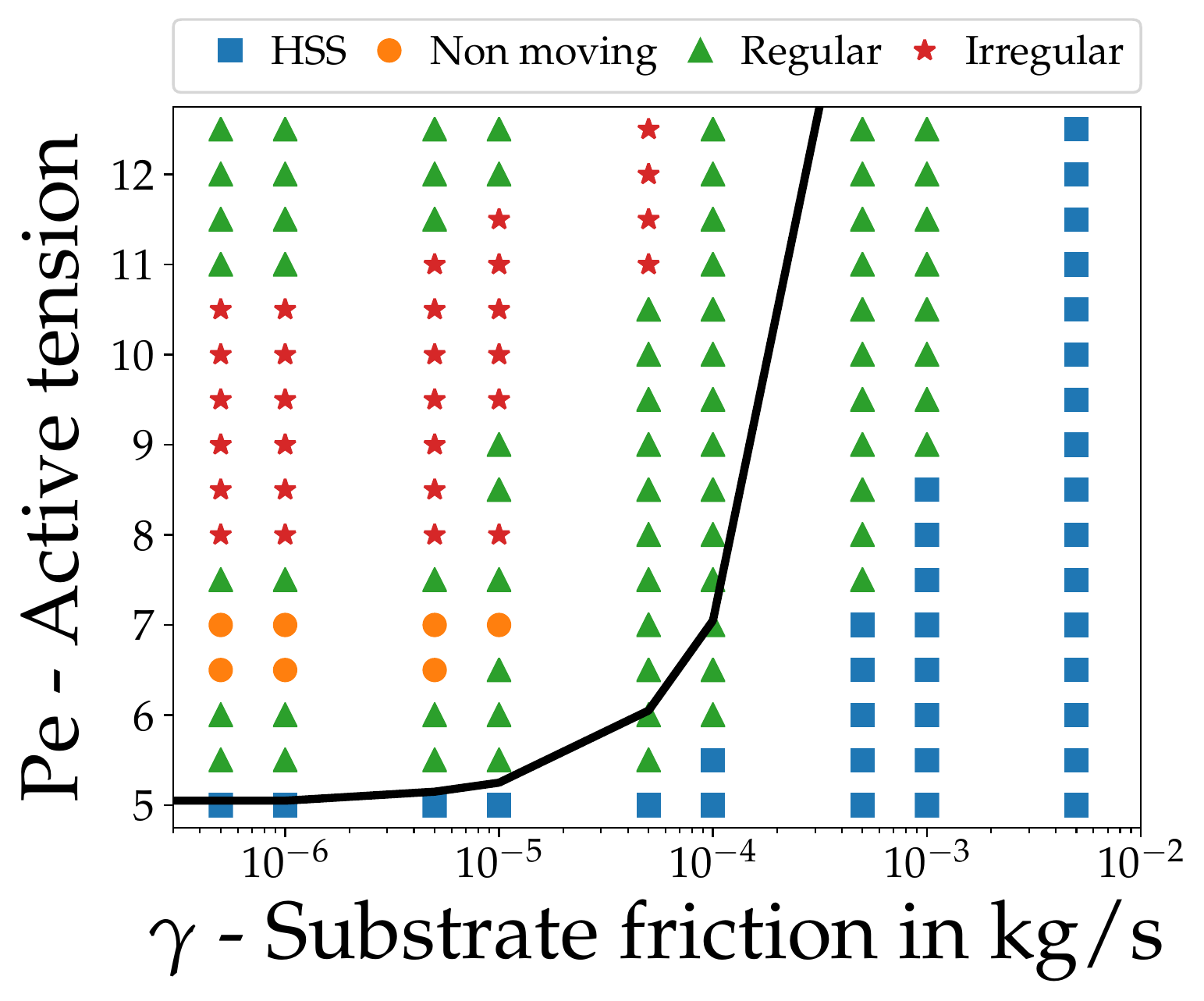}
	\includegraphics[width=0.55\linewidth]{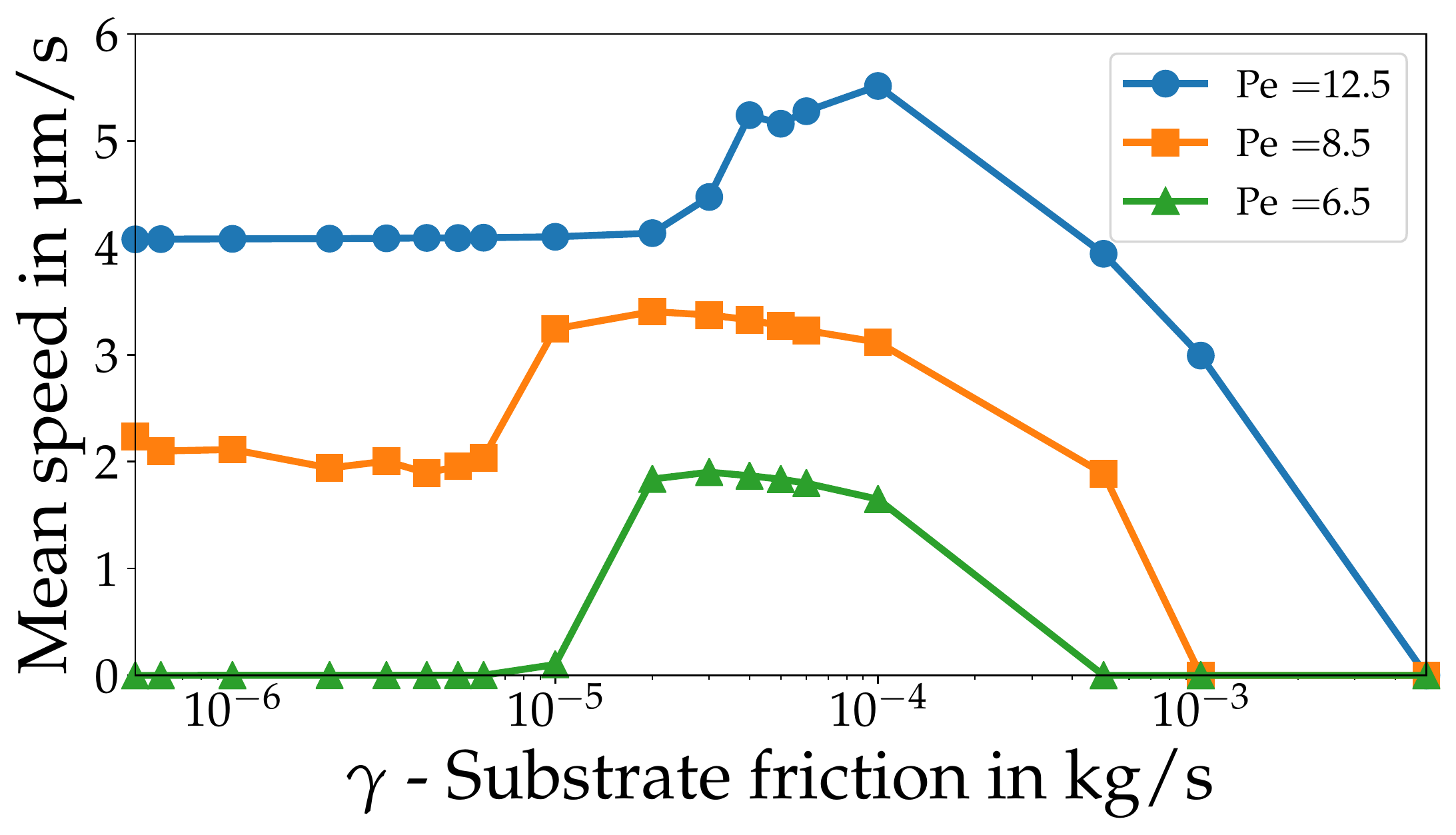}
	\caption{\textbf{Mode of motion (top) and speed (bottom) for different strengths of substrate friction $\boldsymbol{\gamma}$ and active tension as measured by the dimensionless P\'eclet number $\boldsymbol{\mathrm{Pe} = \xi/(D_c\beta)}$.} 
	 Top: We group the results into four different categories: ``HSS'' if the HSS is stable against small perturbations, ``non-moving'', when the HSS is unstable, but the droplet does not move $(K = 0)$,  ``regular'' in the case of periodic movement with less than $5$ Fourier modes $(1 \le K \le 4)$ and ``irregular'' if $K \ge 5$.
	 The thick black line denotes the critical P\'eclet number $\mathrm{Pe}_\mathrm{cr}$ that results from the linear stability analysis.
	 Bottom: The mean droplet speed increases for larger P\'eclet numbers and shows a maximum for medium values of $\gamma$.}
	\label{fig_xi_gamma}
\end{figure}

Depending on the strength of the active tension as measured by the dimensionless P\'eclet number $\mathrm{Pe}$ and the friction between droplet and substrate, the droplet's position is stationary or oscillates with one or more frequencies. We decompose the position over time data in frequency components and use the number of excited Fourier modes $K$ to characterize a droplet's movement. We group the results into four different categories: ``HSS'' if the HSS is stable against small perturbations, ``non-moving'', when the HSS is unstable, but the droplet does not move $(K = 0)$,  ``regular'' in the case of movement with less than $5$ modes $(1 \le K \le 4)$ and ``irregular'' if $K \ge 5$. \figref{fig_xi_gamma} gives an overview which mode of movement emerges (top) and how the droplet speed develops (bottom) when changing active tension strength $\mathrm{Pe}$ and the substrate friction $\gamma$.

In agreement with the linear stability analysis from \cite{radszuweit_intracellular_2013}, the HSS is stable against small perturbations if $\mathrm{Pe}$ is below a critical P\'eclet number $\mathrm{Pe}_\mathrm{cr}$. Above this critical value, the HSS destabilizes and non-linear numerical simulations have to be carried out to determine the resulting patterns. The different categories of movement often arise in bands that appear or disappear when the parameters change. 

The linear stability analysis predicts an oscillatory instability to long-wavelength modes where initially a symmetric standing wave-like pattern with two nodes appears from a superposition of left and right traveling waves. In the long term, however, an antisymmetric standing wave pattern with one node of double wavelength emerges. An adaptation (coarsening) of the pattern towards larger wavelength is often found in such systems. Overall this coarsening is a crucial point, because we show that the change in symmetry of the pattern is what leads to notable motion of the boundary.

If $\gamma$ is too large, the friction inhibits any pattern formation. The HSS is stable, and small perturbations decay. A higher substrate friction shifts $\mathrm{Pe}_\mathrm{cr}$ to higher values, and for $\gamma = \fricU{e-2}$ the HSS is always stable. If $\gamma$ is too low $(< \fricU{e-7})$, there is effectively no friction between droplet and substrate. In this case, our equations do not possess full rank any more and we omit showing these results.

\begin{figure}[!ht]
	\centering
	\includegraphics[width=0.495\textwidth]{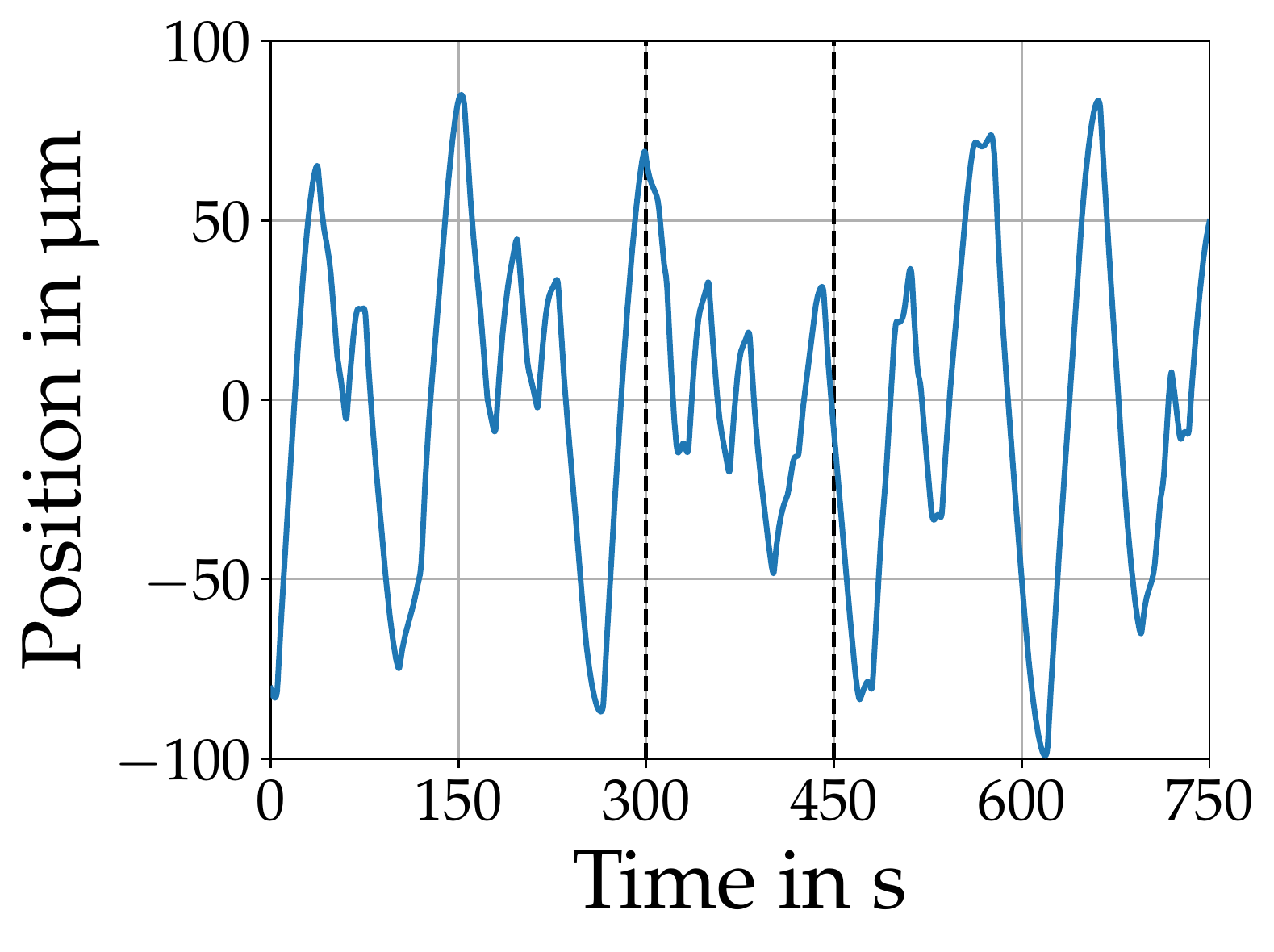}
	\includegraphics[width=0.495\textwidth]{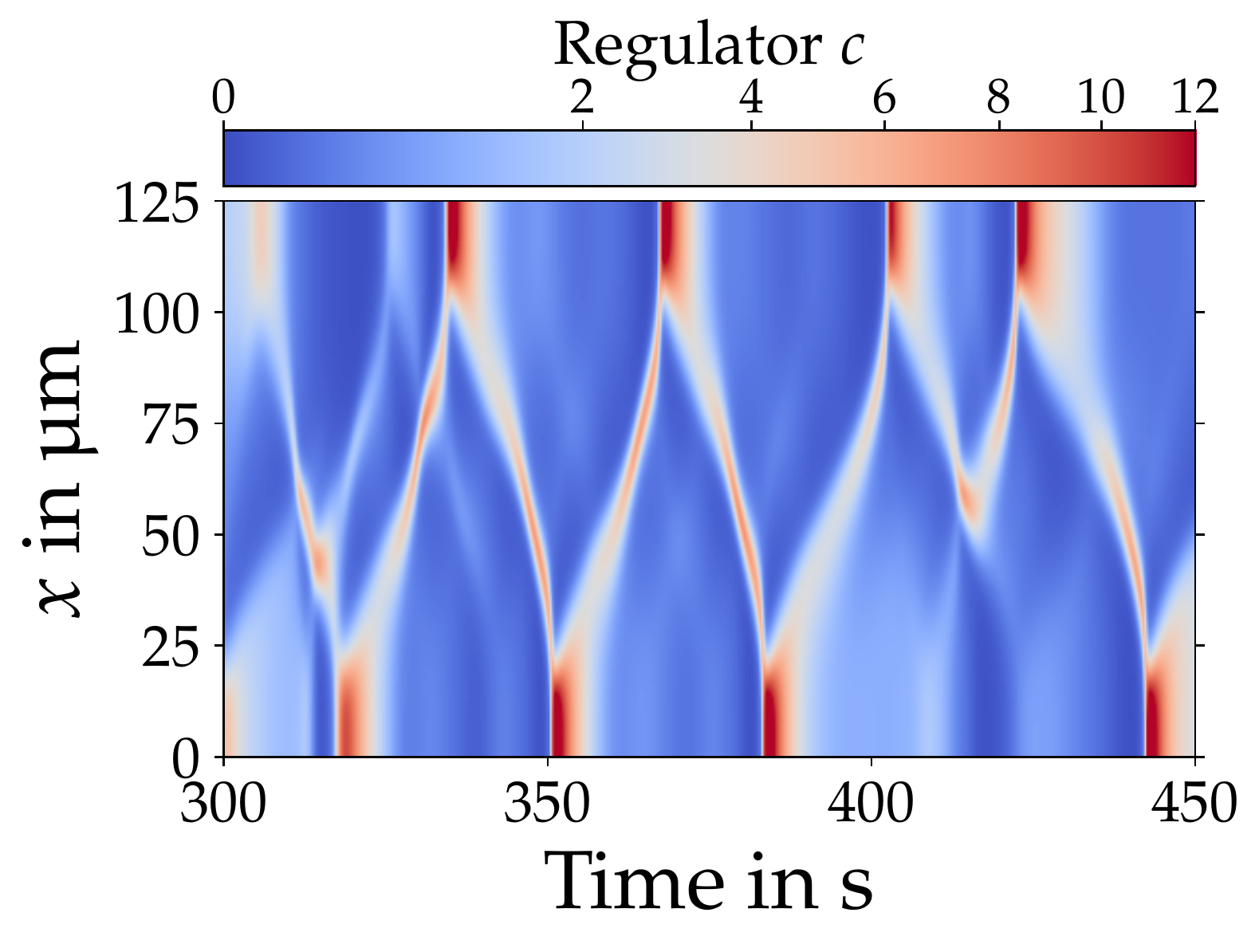}
	\caption{\textbf{Irregular oscillations of the droplet's  position over time (top) appear together with asymmetric regulator profiles (bottom).} With an active tension ($\mathrm{Pe} = 9.5$) higher than in Fig. \ref{fig_xi60} the droplet's movement and the regulator dynamics become irregular and the spectrum of its trajectory is continuous (data not shown).}
	\label{fig_chaos}
\end{figure}

In the region in-between, we can observe regular and irregular oscillatory motion as well as non-moving droplets. For a high substrate friction, the droplet performs regular oscillations. Beginning with $\gamma = \fricU{5e-5}$, irregular oscillations appear under strong enough active tension $(\mathrm{Pe} \ge 11)$. An example is shown in \figref{fig_chaos}. With a further decrease of $\gamma$ this band of irregular solutions is shifted to a lower value of $\mathrm{Pe}$. With $\gamma = \fricU{e-5}$ non-moving solutions occur for intermediate $\mathrm{Pe}$. In contrast to the case of transient symmetric oscillations in \figref{fig_xi60}, here the symmetric oscillations remain stable on the timescale of the simulation length (at least $\s{10000}$).

For $\gamma = \fricU{e-5}$ the critical P\'eclet number is $\mathrm{Pe}_\mathrm{cr} = 5.5$. For $\mathrm{Pe} > 5.5$  the droplet's position exhibits regular oscillations together with antisymmetric regulator oscillations and a period of about $\s{98}$. A further increase yields a regime without motion.  If $\mathrm{Pe} \ge 7.5$, the droplet moves again and the number of excited Fourier modes $K$ rises fast. With a further increase of $\mathrm{Pe}$ the droplet's motion becomes irregular. Between $\mathrm{Pe} = 8.5$ and $\mathrm{Pe} = 9$, the droplet performs a regular motion with a fundamental period of about $\s{94}$. Beginning with $\mathrm{Pe} = 9.5$, the movement becomes irregular again. Upwards from $\mathrm{Pe} = \num{11.5}$ the droplet oscillates regularly, however the amplitude of the fundamental mode has a period of about $\s{25}$ and is much larger than the other ones. Thus, the movement is regular but with a higher frequency than in \figref{fig_xi60}.

The dependence of the mean speed, as measured by the averaged speed of its boundary, displays a maximum for intermediate values of $\gamma$. For $\mathrm{Pe} = 6.5$ the droplet rests for weak and strong friction ($\gamma\le \fricU{e-5}$ and $\gamma\ge \fricU{5e-4}$). In-between it performs regular oscillations with a mean speed of $\velU{2}$. For higher values of $\mathrm{Pe}$, the droplet's speed is shifted to higher values with a maximum speed of about $\velU{3.4}$ for $\mathrm{Pe} = 8.5$ and about $\velU{5.5}$ for $\mathrm{Pe} = 12.5$. Just as for a lower active tension, droplets are at rest for a strong friction and the speed approaches an almost constant value for a weaker friction.

\subsection*{Comparison of our model with experiments on Physarum microplasmodia}
In the following, we compare the predictions of our model with some experimental results recently reported by \cite{zhang_self-organized_2017,lewis_coordination_2015}.
In experiments, directed motion of Physarum microplasmodia was found to be accompanied by oscillations with a period between $\s{85}$ and $\s{110}$. 
During the peristaltic mode of motion, microplasmodia undergo a forward displacement of $d_F \approx\um{40}$, followed by a backward displacement of
$d_B \approx\um{15}$-$\um{20}$. As displayed in \figref{fig_xi_gamma}, we find regular 
oscillations for large parameter regimes in our simulations. As shown in \figref{fig_xi60}, once the dominant pattern with regular, antisymmetric regulator oscillations has emerged, the droplet's position over time oscillates with a period of $T = \s{98}$, which in line with the measurements, and undergoes displacements of 
$d \approx \um{82}$. However, we always find $d = d_F = d_B$ such that no net motion occurs.
\begin{figure}[!ht]
	\centering
	\includegraphics[width=0.5\textwidth]{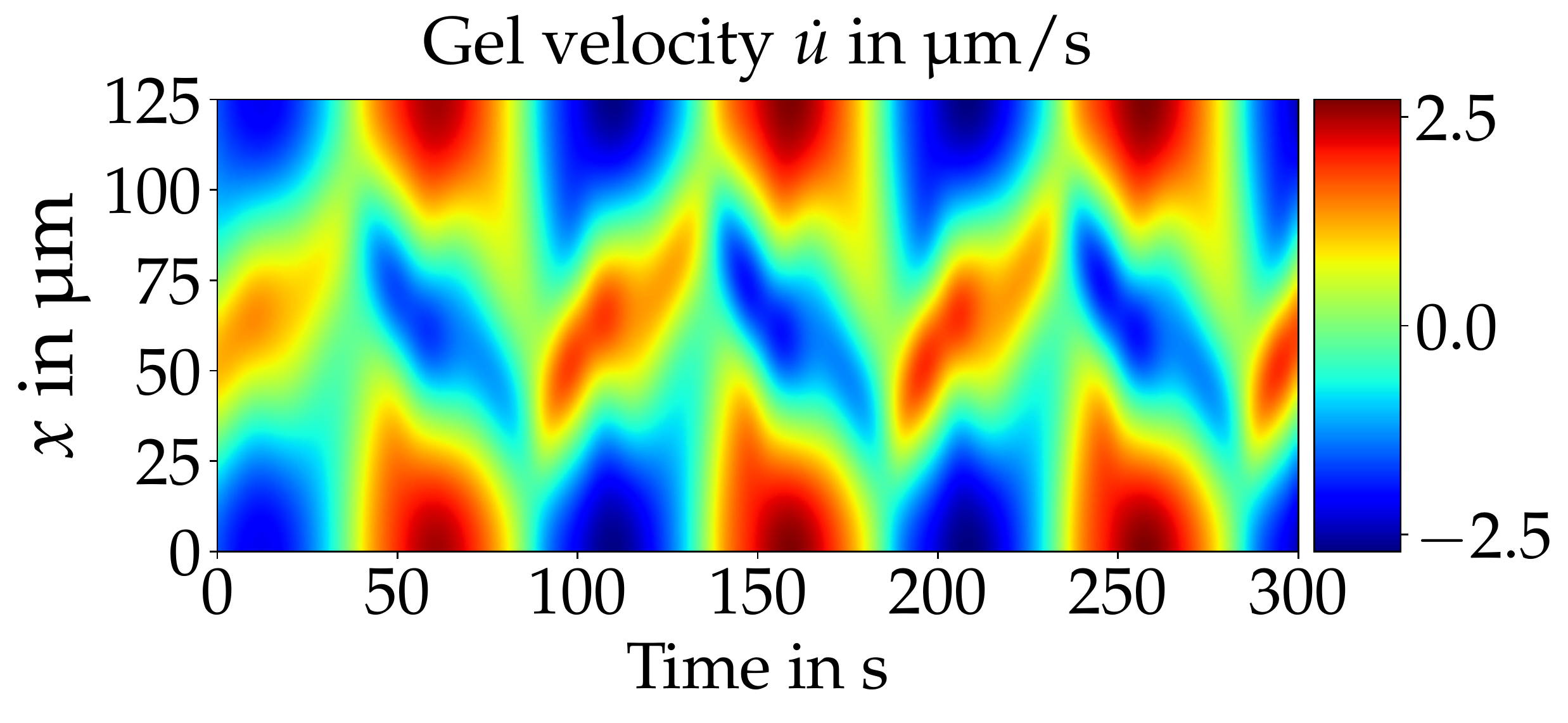}
	\includegraphics[width=0.49\textwidth]{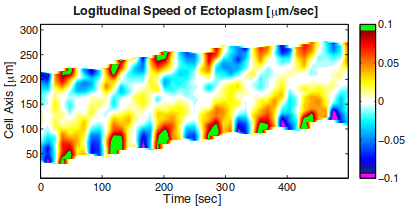}
	\caption{\textbf{Gel velocity in simulation (top) and experiment (bottom).} At a fixed position, the flow direction alternates between forward and 
	backward with a period of about $\s{100}$ in experiment as well as in simulation. Flow at the center is opposite and of weaker magnitude than 
	flow at the boundaries. Experimental backward flow is of weaker magnitude than the forward flow, resulting in a a net motion. Simulated forward and 
	backward flows have equal magnitude and thus cancel exactly.
	Bottom figure taken from \cite{zhang_self-organized_2017}.  Copyright: Journal of Physics D: Applied Physics by IOP Publishing.}
	\label{fig_CompExpMov}
\end{figure}

Additionally, we compare experimentally observed flow patterns within the microplasmodia with our simulations. The space-time plot in \figref{fig_CompExpMov} shows the gel flow (top) and experimentally obtained ectoplasmic flow (bottom) for the peristaltic mode  of motion from \cite{zhang_self-organized_2017}.
At a fixed position in our simulations, the flow alternates periodically between forward and backward flow. While the flow directions at front and back are 
equal, with a higher magnitude at the back, the center part is flowing towards the opposite direction with weaker magnitude than flow at the 
boundaries. 
Experimentally, flows towards the front have a larger magnitude than backward flows, resulting in a net propagation velocity 
of the entire microplasmodia of $v_\mathrm{exp}\approx\num{0.15}$-$\velU{0.2}$. In simulations, forward and backward flows have equal magnitude and thus cancel within one oscillation period.

Spatio-temporal measurements of the regulator dynamics $(Ca^{2+})$ reveal that microplasmodia motility is accompanied by calcium waves. These resemble traveling waves in the peristaltic mode and standing waves in the amphistaltic mode \cite{zhang_self-organized_2017}. In our simulations, the droplets  never exhibit net motion
and we observe only standing or irregular wave patterns. Note, however, that we model the regulator dynamics with an advection-diffusion equation whereas in experimental systems, calcium additionally takes part in chemical reactions \cite{alonso_oscillations_2016,dupont_models_2016}.

%% file: discussion.tex
\section{Discussion}
To model motile cells, continuum mechanical models must be supplemented with free boundary conditions. 
Here, we extended the poroelastic model with rigid boundaries from \cite{radszuweit_active_2014,radszuweit_model_2010,alonso_oscillations_2016,radszuweit_intracellular_2013}
to the case of free boundaries and included linear friction with the substrate. In this minimal model, we explore the conditions for self-organized motion of an active poroelastic droplet. This model is a first step towards a more detailed description of moving Physarum microplasmodia.

We observed different modes of motion ranging from resting droplets to droplets performing regular and irregular oscillations. The symmetry breaking from a standing wave with mirror symmetry and wavelength equal to the system length to an asymmetric standing wave with half a wavelength in the system after a long transient is necessary for motion of the boundary (\figref{fig_xi60} and \figref{fig_xi60Snap}). In \cite{radszuweit_intracellular_2013}, only symmetric oscillations were reported, and there were no transitions to asymmetric oscillations. In addition, we found that the droplet's speed has a peak for intermediate values of $\gamma$ (\figref{fig_xi_gamma}).

In our earlier work \cite{radszuweit_intracellular_2013}, traveling wave patterns appeared that were reflected at the boundaries and moved back and forth. The chaotic pattern is an intermediate state between the region of stable standing waves (symmetric or asymmetric) and the traveling domain pattern. The different patterns have mostly physiological significance as the quantitative and qualitative aspects of droplet motion in the model change.

In the parameter plane spanned by the friction coefficient ($\gamma$) and the strength of active tension ($\mathrm{Pe}$), we identified parameters that reproduce experimentally observed oscillation periods of about $\s{100}$ \cite{zhang_self-organized_2017,lewis_coordination_2015}. 
Additionally, as shown in \figref{fig_CompExpMov}, the simulated flow patterns are qualitative in line with experimentally measured flow patterns.

The obvious question arises why, for all cases of periodic and even irregular motion, the time averaged position vanishes, and no net motion occurs. An essential ingredient to achieve directed motion is a mechanism which breaks the front-back symmetry and thus establishes a polarity \cite{rappel_mechanisms_2017}. 
The regulator distributions produced by pure diffusion-advection dynamics may look asymmetric at certain instants in time, but the long time averaged distribution is always symmetric. While this is expected for a regular oscillation such as \figref{fig_xi60}, it is surprising for the irregular dynamics as shown in \figref{fig_chaos}. 
The absence of a time-averaged asymmetry in the regulator dynamics for the irregular case indicates that an additional mechanism to establish a polarity is required to model the experimentally observed motion of Physarum microplasmodia.

In \cite{lewis_coordination_2015,lewis_analysis_2017}, the front-to-back symmetry was broken by introducing externally imposed traveling waves of friction strength and contractile stress.
Another approach to break the spatial symmetry in the time averaged distribution of the regulator variable $c$ is to include reaction kinetics for the regulator as done in models for resting Physarum droplets earlier \cite{radszuweit_active_2014,alonso_oscillations_2016}. There, unidirectional traveling mechano-chemical waves were reported in contrast to the back and forth moving waves found in our study here and previously in \cite{radszuweit_intracellular_2013}. The reaction kinetics allows for a temporal variation of the total concentration of the regulator encoded in the variable $c$. Additionally, the total amount of the regulator is not conserved anymore.

Moreover, the following argument shows that net motion 
is in general impossible for a substrate friction $\gamma$ constant in space.
For constant mass density, the velocity $\bar{v}$ of the droplet's center of mass is obtained by spatially averaging the gel velocity
\begin{eqnarray}\label{eq_AvgVel}
\bar{v} = \frac{1}{V_0}\int_\mathcal{B}\dot{u}dx,
\end{eqnarray}
where $V_0$ denotes the constant volume of the droplet (length in one spatial dimension).
Adding the force balances for gel and fluid phases, \eq{eq_ForceBal_1} and \eq{eq_ForceBal_2}, yields
\begin{eqnarray}
\bar{v} = \frac{1}{V_0}\int_\mathcal{B} \frac{1}{\gamma}\partial_x\left( \sigma -p\right) dx.
\end{eqnarray}
For $\gamma$ constant in space, we may use the Gauss theorem to transform the volume integral to a surface integral with normal vector $n$,
\begin{eqnarray}\label{eq_NoNetMotion}
\bar{v} = \frac{1}{\gamma V_0} \oint_{\partial\mathcal{B}}\left( \sigma -p \right) n dS.
\end{eqnarray} 
Together with the free boundary condition \eq{eq_BCstress}, we immediately obtain a vanishing center of mass velocity $\bar{v} = 0$.
This is confirmed by a direct calculation of $\bar{v}$ in numerical simulations as given by \eq{eq_AvgVel}.

Thus, a spatially dependent substrate friction seems to be an essential ingredient to obtain net motion. 
This is in line with the results from \cite{lewis_coordination_2015}, where a space-dependent friction coefficient was introduced and net motion was observed. 
In addition, the authors in \cite{zhang_self-organized_2017} found evidence for a nonlinear relationship between microplasmodia velocity and traction force, yielding a position-dependent friction coefficient. 

A recent study suggests that the activity difference between the two phases can lead to a phase separation \cite{weber_differential-activity_2017}. A future extension of the presented model may include this phenomenon by treating the phase composition as a spatially dependent local variable.

With minor modifications, our model might be applicable to other systems. We highlight photosensitive self-oscillating gels where the swelling is regulated by an embedded, light-sensitive chemical reaction \cite{lu_photophobic_2013,ren_retrograde_2016, ren_autonomous_2017,epstein_photo-controlled_2017}. Applying appropriate illumination pattern 
generates directed waves of a regulator which lead to oscillatory motion of the gel with \cite{lu_photophobic_2013} and without \cite{ren_autonomous_2017} 
net motion. Some parameters in this chemical reaction are easier to control experimentally than in Physarum microplasmodia. Therefore, it could be a helpful setup to study the transition from motion without to motion with net motion.

%% file: method.tex
\section*{Appendix A: Numerical Details}
We solve the equations of motion on an one-dimensional Chebyshev-Lobatto grid \cite{trefethen_spectral_2000} of size $L$ with $N$ points. We utilize no-flux boundary conditions for the concentrations $c$ and free boundary conditions for the mechanical equations. We formulate our model in the gel's body reference frame (BRF). For details about the derivation of the model refer to \cite{radszuweit_active_2014,alonso_oscillations_2016}

We split the full equations from from \eq{eq_FullEq_a} - \eq{eq_FullEq_d} into a mechanical and an advection-diffusion part and solve each part separately. We use pseudo-spectral methods (Chebyshev) for the discretized spatial derivatives and the Euler method for time-stepping.

For the mechanical part, we introduce $U\equiv\partial_t u = \frac{u-u^t}{\Delta t}$, where $t$ denotes the current time-step and variables without explicit time dependency are at time $t+\Delta t$. Then, we arrive at
\begin{eqnarray}
u - U\Delta t &= u^{t}\\
\rho_f\eta_f\partial_{xx} v + \rho_g\eta_g\partial_{xx} U + \rho_g E\partial_{xx}u\nonumber\\
   - \gamma\rho_g U - \partial_x p &= -\rho_g\partial_x T(c^{t})\\
\eta_f\partial_{xx} v - \rho_g\beta(v-U) - \partial_x p &=0\\
\partial_x (\rho_g U + \rho_f v) &= 0.
\end{eqnarray}
\noindent
Then, we solve the advection-diffusion part semi-implicitly	with $\partial_t c = \frac{c-c^t}{\Delta t}$. This approach yields
\begin{eqnarray}
c &+\Delta t \partial_x\left(w^t c\right) - \Delta t D_c \partial_{xx} c = c^{t},
\end{eqnarray}
where $w= v-U$ is the fluid velocity in the gel's BRF. This yields the linear equation
\begin{eqnarray}
\left(\mathds{1} - \Delta t D_c \partial_{xx} + \Delta t \partial_x w^t \right) c = c^{t}.
\end{eqnarray}

\noindent
We solve our equations using python \cite{oliphant_python_2007} with the iterative gmres solver from scipy and an ILU preconditioner.

\section*{Appendix B: Regulator concentration in body reference and lab frame}\label{sec_BodyVsLab}
\begin{figure}[!ht]
	\centering
	\includegraphics[width=0.7\linewidth]{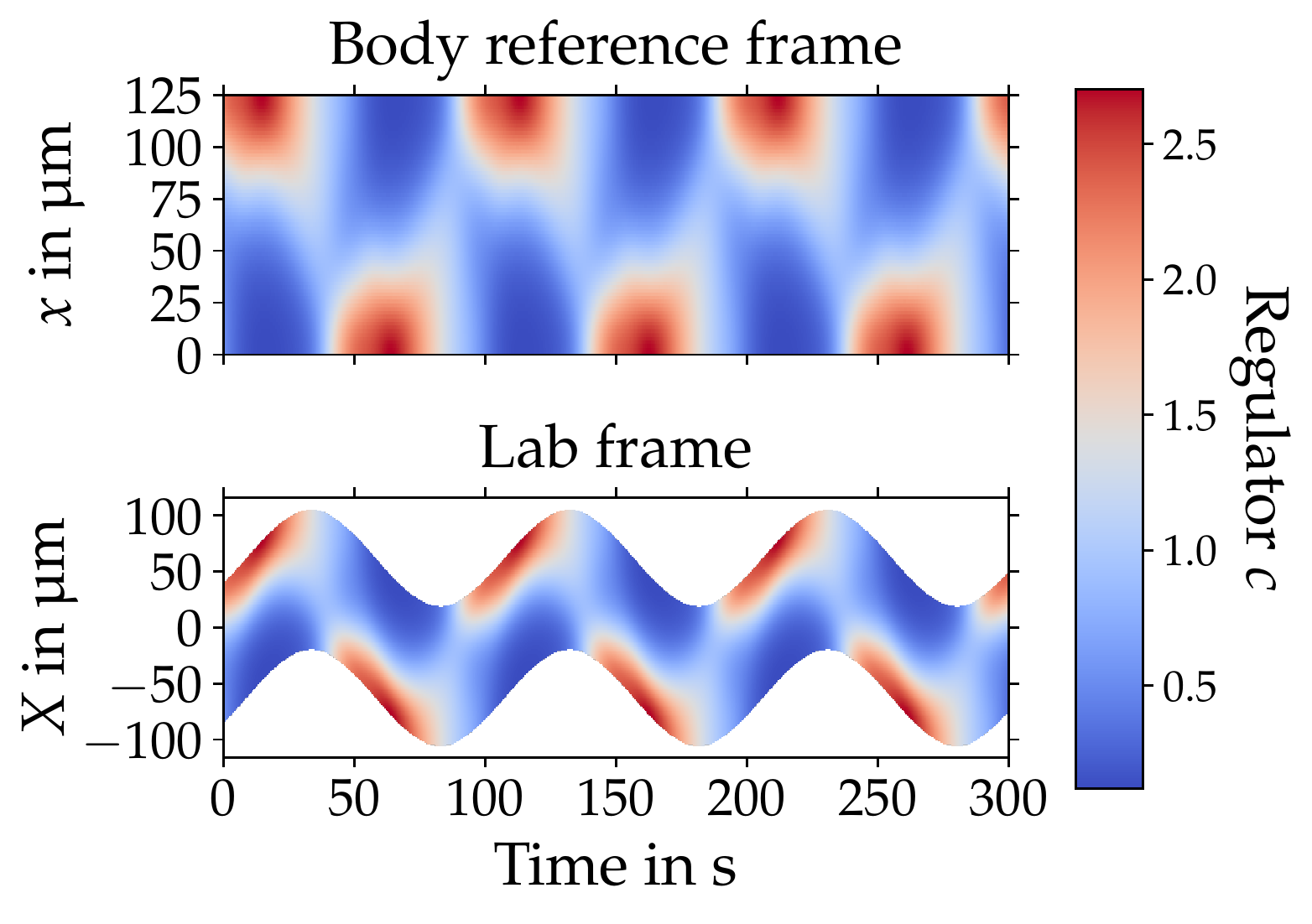}
	\caption{ \textbf{Spatially antisymmetric regulator oscillation in body reference (top) and lab frame (LF) (bottom).} In the LF, the droplet is moving into the direction where $c$ has a local maximum. The regular regulator oscillation yields a periodic movement with a period of about $\s{98}$.}
	\label{fig_BRF}
\end{figure}

We solve our model equations in the gel's BRF and the resulting quantities are defined in this frame. However, we as observers are located in the lab frame (LF). \figref{fig_BRF} shows how a regular regulator oscillation (compare with Fig. 1 in the main text) looks in BRF as well as LF. The transformation of quantities from the BRF to the LF is given by the deformation field $u$ with $X_0 = x_0 + u(x_0)$, where $x_0$ the position in the BRF and $X_0$ is the position in the LF.

The concentration in \figref{fig_BRF} exhibits spatially antisymmetric oscillations in the BRF and switches between a state with a high value of regulator at the left boundary and a low value at the right boundary and the reversed state. In the LF, this yields a periodic movement of the droplet. As long as the concentration has a local maximum at a certain boundary the droplet is moving into this direction.

\section*{Appendix C: Parameters}

\begin{table}[H]
	\begin{center}
		\caption{ \textbf{Model parameters}} 
		\begin{tabular}{|lp{4.0cm}ll|}
			\hline
			Par & Description & Value & Units \\
			\hline
			$N$ & Number of grid points & $120$ &  -\\
			$\Delta t$ & Numerical time step & $0.001$ &  $\si{\s}$\\
			$D_c $ & Regulator diffusion & $200$ &  $\si{\square\micro\meter\per\s}$\\
			$L $ & Length &  $125$  & $\si{\micro\meter}$  \\
			$\rho_g$ & Gel fraction & $0.5$ & -\\
			$\rho_f$ & Fluid fraction & $0.5$  & -\\
			$\eta_g$ & Viscosity gel& $10^{-2}$  &  $\si[per-mode=fraction]{\kg\per\micro\meter\per\s}$\\
			$\eta_f$ & Viscosity fluid& $2\times 10^{-8}$ & $\si[per-mode=fraction]{\kg\per\micro\meter\per\s}$ \\
			$\beta$ & Friction between both phases & $10^{-4}$ & $\si[per-mode=fraction]{\kg\per\micro\meter\cubed\per\s}$\\
			$E$ & Young modulus & $0.01$ &$\si[per-mode=fraction]{\kg\per\micro\meter\per\square\s}$ \\
			$\mathrm{Pe}$ & Active tension & $6$ & - \\
			$\gamma$ & Substrate friction & $10^{-5}$  &$\si{\kg\per\s}$ \\
			\hline
		\end{tabular}
		\newline
		\newline
		These parameters are used throughout this work any deviation is explicitly marked. These parameters are based on typical estimates for eukaryotic cells. Taken from \cite{radszuweit_intracellular_2013}.\label{tab_param}
\end{center}
\end{table}